\documentclass[manuscript,screen]{acmart}
\AtBeginDocument{%
  \providecommand\BibTeX{{%
    \normalfont B\kern-0.5em{\scshape i\kern-0.25em b}\kern-0.8em\TeX}}}

\setcopyright{acmcopyright}
\copyrightyear{2022}
\acmYear{2022}
\acmDOI{10.1145/1122445.1122456}

\acmJournal{JACM}
\acmVolume{37}
\acmNumber{4}
\acmArticle{111}
\acmMonth{8}

\copyrightyear{2022} 
\acmYear{2022} 
\setcopyright{acmlicensed}\acmConference[FAccT '22]{2022 ACM Conference on Fairness, Accountability, and Transparency}{June 21--24, 2022}{Seoul, Republic of Korea}
\acmBooktitle{2022 ACM Conference on Fairness, Accountability, and Transparency (FAccT '22), June 21--24, 2022, Seoul, Republic of Korea}
\acmPrice{15.00}
\acmDOI{10.1145/3531146.3533190}
\acmISBN{978-1-4503-9352-2/22/06}

\begin{document}

\title{The Case for a Legal Compliance API for the Enforcement of the EU's Digital Services Act on Social Media Platforms}
\author{Catalina Goanta}
\email{e.c.goanta@uu.nl}
\orcid{0000-0002-1044-9800}
\affiliation{%
  \institution{Utrecht University}
  \city{Utrecht}
  \country{The Netherlands}
}
\author{Thales Bertaglia}
\orcid{0000-0003-0897-4005}
\email{t.costabertaglia@maastrichtuniversity.nl}
\author{Adriana Iamnitchi}
\orcid{0000-0002-2397-8963}
\email{a.iamnitchi@maastrichtuniversity.nl}
\affiliation{%
  \institution{Maastricht University}
  \city{Maastricht}
  \country{The Netherlands}
}


\renewcommand{\shortauthors}{Goanta et al.}

\begin{abstract}

  In the course of under a year, the European Commission has launched some of the most important regulatory proposals to date on platform governance. 
  The Commission's goals behind cross-sectoral regulation of this sort include the protection of markets and democracies alike. While all these acts propose sophisticated rules for setting up new enforcement institutions and procedures, one aspect remains highly unclear: how digital enforcement will actually take place in practice. Focusing on the Digital Services Act (DSA), this discussion paper critically addresses issues around social media data access for the purpose of digital enforcement and proposes 
  the use of a legal compliance application programming interface (API) as a means to facilitate compliance with the DSA and complementary European and national regulation. To contextualize this discussion, the paper pursues two scenarios that exemplify the harms arising out of content monetization affecting a particularly vulnerable category of social media users: children. 
  The two scenarios are used to further reflect upon essential issues surrounding data access and legal compliance with the DSA and further applicable legal standards in the field of labour and consumer law.

\end{abstract}




\maketitle





\section{Introduction}

In the European Union, platform governance regulation is in full swing. In the course of under a year, the European Commission has launched some of the most important legal proposals to date with respect to harms arising in the digital economy: the Digital Services Act (DSA) package \cite{noauthor_digital_nodate}, the Artificial Intelligence (AI) Act \cite{noauthor_eur-lex_nodate}, and the regulatory package relating to political advertising and disinformation \cite{noauthor_press_nodate}, to name a few. The Commission's goals behind cross-sectoral regulation of such sort include the protection of markets and democracies alike \cite{noauthor_digital_nodate, cauffman_goanta_2021}. 

This paper focuses on the DSA and particularly on how it is supposed to facilitate digital enforcement \cite{Peukert2022, Duivenvoorde2022}. In spite of being a comprehensive proposal that tackles both substantive legal standards and proposes new procedures and institutions, the technical dimension of digital enforcement, namely how the compliance with legal obligations will be observed in practice, remains underdeveloped. According to the DSA proposal, this will be outsourced to delegated acts (Article 31(5) DSA), which are additional legal standards prepared and adopted by the European Commission, aimed to achieve uniformity after consultations with expert groups. 

The legal compliance of platforms with existing legislative frameworks is a complex discussion spanning across a wide range of legal fields and regulatory sectors. This is particularly the case for social media platforms, which are undertaking a visible shift in business models. Traditionally, social media has been seen as a space of user-generated content, and much attention has been drawn to harms associated with this sharing, such as disinformation~\cite{hills_information_nodate}, incitement to violence/hate speech~\cite{sigurbergsson-derczynski-2020-offensive, bertaglia-etal-2021-abusive}, and promotion of unhealthy behavior for young adults~\cite{safi-samghabadi-etal-2020-detecting}. Yet little acknowledgement has been given to trends that are significantly changing the social media landscape, such as content monetization. 

Monetization, or simply put, the ability to make money out of content posted online \cite{noauthor_monetisation_nodate}, has a wide array of variations, ranging from advertising to the sale of goods/services and crowdfunding. At the heart of the monetization ecosystem on social media are influencers (also known as content creators)\cite{10.1145/3359321}, who engage in complex supply chains for the pursuit of revenue based on their Internet activities. As a pervasive type of Internet entrepreneurs engaged in cultural production, influencers often turn into media empires with impressive economic returns \cite{reddy-etal-2021-detecting}. With the content creation economy booming globally and at unprecedented scale, public authorities tasked with the enforcement of existing Internet law (e.g. media, consumer protection) are often infrastructurally overwhelmed. Yet in the coming years, for legal compliance on digital markets to be in any way effective, it will need to be scalable.  


Taking the DSA as an example, this discussion paper critically reflects on the needs for standardization and proposes a conceptual solutions for how data sharing ought to be done by digital platforms in compliance with the DSA. To contextualize these needs, we look at two scenarios of harms arising out of content monetization. Monetization has been fueling the creator economy and in doing so, has been reconfiguring the stakeholders of the social media ecosystem and their commercial interests. Within this ecosystem, some creators and some parts of the audience are more vulnerable than others: for example, children. While legal compliance is relevant for content monetization as a whole, this paper focuses on compliance with rules that are relevant for children involved in the content creation economy either as influencers or audiences, given their particular need for protection. For terminological purposes, we must clarify that although conceptual differences may exist, and more specific definitions may be employed in procedural law, we use the terms legal compliance and legal enforcement interchangeably to mean the process of ensuring that legal rules are applied by those bound by them in practice.    

Legal compliance, even when the rationale for protecting individuals is so societally necessary as in the case of children, comes with tremendous hurdles. Social media platforms control the activity of their users, and collect all the data they produce \cite{goantaOrtolani2021}. This puts platforms in the unique position of monitoring legal compliance, and designing processes to facilitate this legal goal. However, commercial interests generally compete with the overwhelming complexity of legal systems characterized by overlaping and sometimes conflicting legal fields, jurisdictions and practices. As a result, social media platforms set their own private standards, and generally ignore national or supranational law unless there are serious risks for legal enforcement. A concrete example is the General Data Protection Regulation (GDPR) \cite{noauthor_eur-lex_nodate-1}, a legal instrument so threatening from the perspective of legal compliance, that it has led to the adoption of separate terms of service by platforms to display to EU IP addresses. Another example is the Digital Millenium Copyright Act \cite{noauthor_digital_nodate-1}, which heightened penalties for copyright infringements on the Internet, and which has shaped the \textit{sui generis} dispute resolution mechanisms that platforms developed in order to avoid legal liability for copyright infringements. 

Such examples show how legislative reforms shaped legal compliance measures taken by social media platforms. However, this is not the case for all legal standards applicable to online content, where incentives to comply with the law may be lower. Media law, consumer protection or labour law authorities just to name a few, have only recently started developing approaches, both manual and automated, for the monitoring of social media content with the purpose of checking for legal compliance. The sheer scale of social media platforms and generally the information asymmetry between tech companies and public authorities make it virtually impossible for legal compliance verification to be systematic and comprehensive. On the one hand, authorities do not have access to the same data as platforms. In some fields, like consumer protection, the powers of authorities in the field of market surveillance have been considerably widened. This means generous amounts of data can be requested in the course of investigations regarding consumer violations. However, what data is necessary and whether the requested data is not tampered with are questions that remain open both from a legal and technical perspective. On the other hand, even if relevant and accurate data is successfully retrieved from platforms, data analysis as a computational burden still lies with authorities in proving regulatory violations. As the European legislator and national authorities increasingly rely on digital monitoring and enforcement, important questions arise relating to how platform data sharing can be standardized for the operationalization of legal compliance \cite{euwhitepaper2020, schrepel2021computational}. It is worth noting that in this paper we use 'legal compliance' and 'digital enforcement' interchangeably, as they can be seen as different sides of the same phenomenon: that companies have a duty to comply with the law, and national authorities have the legitimacy to pursue the application of the law in practice. 




The paper is structured as follows. The first part describes two different online harm scenarios arising on social media platforms as a result of content monetization and discusses them from the perspective of applicable European or national law as content regulation. The second part reflects upon how legal enforcement can be translated into computational tasks, and what data is necessary for such tasks. The third part proposes an original solution for digital enforcement in the form of a legal compliance API, and addresses some of its general characteristics, benefits and challenges.

\section{Two harms scenarios and compliance reflections}
\label{sec:scenarios}

The regulation of online harms is an ongoing policy concern for law-makers around the world \cite{Gorwa2019, Rieder2020}. In the EU, the DSA aims to establish a new legal regime for platform liability and transparency with respect to illegal content \cite{Duivenvoorde2022, Peukert2022}. Generally, illegal content has been used to refer to the proliferation of content causing extreme categories of harm, such as child pornography or terrorism. However, those conceptions only partially define the wide concept of illegal content. The DSA refers to it as ``any information, which, in itself or by its reference to an activity, including the sale of products or provision of services is not in compliance with Union law or the law of a Member State, irrespective of the precise subject matter or nature of that law'' (Article 2(g)). It follows that the illegality of content is dependent on the standards of legal compliance in sectors which are relevant for digital services. With the DSA's focus on transparency in compliance and digital monitoring set out through new procedural obligations, it is more important than ever to acknowledge that any field of European (or national) law can be a source of content regulation. For example, not only criminal law and anti-discrimination law are relevant for content posted online because they may shape the legal definitions of concepts such as incitement to hatred or hate speech. Consumer protection law is a field of law equally important to content regulation, as many others. Yet establishing what regulation governs social media content can be a gargantuan task. Factors such as fast-paced digital business models and legislative inflation have fuelled a web of rules which is often difficult (if not impossible) to navigate. 

This section proposes a legal mapping exercise by focusing on two scenarios of harms relating to the activity of children on social media. These scenarios have been selected for two main reasons. 
First, they portray different types of harms arising out of content monetization. As one of the main trends driving new social media products, content monetization creates new participation incentives in the platform economy, especially for peer entrepreneurs (e.g. freelancers), and especially if such peer entrepreneurs are minors or affect minors as audiences. 
Second, 
they showcase the breadth of content regulation when it comes to the assessment of harms. Fundamental rights, such as the freedom of expression, are only part of the narrative around illegal content and online harms. The two scenarios address legal considerations relating to illegal content pertaining to labour law, media law and consumer protection, as sources of content regulation which triggers platform liability under DSA compliance mechanisms. Furthermore, the scenarios showcase recent regulatory developments at European and national level which establish clear legal standards. It is worth noting that this analysis does not aim to exhaustively map applicable legal regimes, but rather to exemplify the data issues around legal compliance by selecting and discussing specific provisions from national or European legal instruments.

\subsection{Scenario 1: Child influencers and content as labour}

According to Socialblade, one of the leading social media analytics platforms globally, out of the most viewed 10 YouTube channels, five are channels for children, of which three feature child influencers as the main content creators. In a lot of jurisdictions around the world, child labour under a certain age is prohibited, in order to protect the well-being of minors and ensure their participation in education. The law allows for exceptions to these harsh rules in the form of administrative permits which guardians of children can apply for in situations where children engage in entertainment activities. These rules have been generally adopted as a result of the rise of legacy media during the 20th century, when a high demand for child performers was created. Yet more recently, with the exponential growth of social media platforms, content made by child influencers is becoming increasingly accessible and profitable. Given the popularity of a brand based on the image and reputation of a minor, questions arise relating to what is permissible in terms of the time spent recording, the potential psychological harms of dealing with viral popularity, as well as potential violations of mandatory laws relating to schooling. We call child influencer accounts the accounts that address a child audience and are monetised, therefore independent of the success of the account in terms of number of viewers/followers. For this scenario, the focus is on child labour protections reflected by national law, in particular French law.

While many jurisdictions struggle to understand this phenomenon in order to potentially regulate it, France recently amended its Labour Code (\textit{LOI n° 2020-1266 du 19 octobre 2020 visant à encadrer l'exploitation commerciale de l'image d'enfants de moins de seize ans sur les plateformes en ligne}) to govern the economic activities of influencers below the age of 16 years. It is the only jurisdiction which has adopted legislative amendments to its labour standards as a reaction to the controversial topic of digital work done by children. The law reflects eight articles setting up a new legal regime for activities undertaken 'by an employer engaged in the business of making audiovisual recordings in which the principal subject is a child under the age of sixteen, for distribution for profit on a video sharing platform service' (Article 1.4.b). Child influencer activities that entail a direct or indirect revenue are registered with authorities and can only be performed if there is a permit issued by said authorities: 'the dissemination of the image of a child under sixteen years of age on a video sharing platform service, when the child is the main subject, is subject to a declaration to the competent authority by the legal representatives' (Art. 3.I). 

Authorities currently do not monitor social media at scale, and rely on complaints and small investigations (e.g. focused on specific child influencers) to understand the digital ecosystem. At the time of writing this paper, no platform offers the possibility to link administrative obligations users may have under e.g. labour law, to the accounts that enable social media activities for said users. While it is true that an obligation for obtaining an administrative permit in the case of child influencers is at this moment particular to French law, there is no computational framework for the legal compliance of this obligation. 

Two other legal obligations in the Labour Code modifications include the obligation incumbent on the social media platform 'to promote the reporting, by their users, of audiovisual content featuring children under sixteen years of age that would undermine their dignity or moral or physical integrity' (Article 4.3), as well as 'to facilitate the implementation, by minors, of the right to the deletion of personal data [...] and to inform them, in clear and precise terms, easily understandable by them, of the modalities of implementation of this right' (Article 4.6).


\subsection{Scenario 2: Disclosing advertising on child accounts} 

Influencer marketing is a form of native advertising on social media which has been on the rise for the past years. Generally based on electronic word-of-mouth (e-WOM) \cite{Park_2012}, influencer marketing entails the promotion of goods and/or services by social media influencers, who in turn receive direct (money) or indirect revenue (goods and/or services) from the brands they promote. Influencer marketing variations include:
\begin{itemize}
    \item endorsements - influencers are brand ambassadors who receive money in order to promote goods or services for brands \cite{james2017}.
    \item barter - influencers exchange advertising services against goods or services offered by the brands they promote (e.g. hotel rooms). 
    \item affiliate marketing - influencers receive a commission for every sale concluded by the brand using the influencer's services, often traceable via discount codes \cite{2018}. 
\end{itemize}
This form of monetizing social media content has nurtured a competitive and ever-growing market of 'human ads' who must keep up with fast-changing platform affordances and find clients among brands looking for advertising services, all while making sure that their follower base is entertained, engaged, and ready to be influenced. In marketing, influence often translates into converting ads into sales, a path that is becoming increasingly shorter with the rise of social commerce functionalities (e.g. the Checkout feature on Instagram). 

The popularity of influencer marketing has normalized it as a go-to marketing approach for social media advertising. Yet factors such as low legal information literacy, the attempt to game seemingly biased platform technologies (e.g. recommender systems) or the mere impression of operating in a grey legal area generally drive influencers to hide advertising. Inconspicuous advertising can be achieved in different ways, such as not disclosing to audiences when an influencer is paid to use/review/advertise a product or a service; being sent goods/freebies and positively reviewing them although the personal opinion of the influencer advertising the good differs.

Inconspicuous advertising is a known consumer harm, as it creates situations wherein consumers can be manipulated into transactions they would otherwise not engage in \cite{pottinger2018}. This risk is particularly high with children, whose credulity make them more vulnerable to manipulation \cite{campbell2016}. Advertising that is native to entertainment content is not a new phenomenon (e.g. product placement in movies). However, two complications arise with its newest social media iteration. First, product placement and other forms of native advertising in legacy media (e.g. radio, tv, printed press) have been intensely regulated at the end of the 20th century across the globe, leading to media practices that made advertising recognizable and distinct from other media content. With the rise of native advertising on social media, regulatory questions should not focus on reinventing the wheel, in so far as solutions to this phenomenon already exist, but rather understand how to better tailor current standards for new media contexts. Second, native advertising on social media may pose additional harms in terms of the relatability and accessibility of influencers as idols, with whom audiences develop parasocial relationships. It is thus all the more important to separate commercial interests from socio-cultural experiences underlying social media interactions. 

At European level, the Unfair Commercial Practices Directive (UCPD) is the instrument that most directly covers the phenomenon of surreptitious advertising \cite{purnhagen_2017}. The UCPD is a so-called maximum harmonization instrument, meaning that Member States need to apply the same standards when transposing it at national level. One of the features of the maximum harmonization proposed by the UCPD has been the Annex of practices considered unfair under any circumstance, and are thus prohibited. Point 11 of the Annex is particularly relevant, since it deals with earlier media practices around hidden advertising (e.g. advertorials): 'using editorial content in the media to promote a product where a trader has paid for the promotion without making that clear in the content or by images or sounds clearly identifiable by the consumer'. Recent case law from the Court of Justice of the European Union (CJEU) has further clarified the ambit of this provision. In \textit{Peek \& Cloppenburg} \cite{noauthor_eur-lex_nodate-2}, the Court held that point 11 of the Annex 'was designed, inter alia, to guarantee that any publication over which the trader concerned has exercised influence in its commercial interests is indicated clearly and is understood as such by the consumer'. In other words, any media content that broke the editorial veil of neutrality of a publication using incentives such as payment or other benefits for the author of the content, must be disclosed as such to consumers. If it is not disclosed, it is an unfair practice. The relevance of \textit{Peek \& Cloppenburg} goes even further, as influencer marketing practices have often differentiated between sponsorships (e.g. paying for influencer marketing), on the disclosure of which there was an early industry-level awareness, and barters, where goods and services are exchanged. In the latter case, the influencer industry has been more reluctant to embrace disclosures on the basis of the argument that there is no payment made. This was exactly the problem at the heart of the judgment in \textit{Peek \& Cloppenburg}, and the Court stated that 'to interpret the concept of ‘payment’, within the meaning of [point 11], as meaning that it requires the payment of a sum of money does not reflect the reality of journalistic and advertising practice and would in large measure deprive that provision of its effectiveness'.


\section{Critical reflections on digital enforcement and data access}

\subsection{Barriers to digital enforcement}

Digital enforcement has been traditionally referred to as the process through which rules initially intended for the offline world are enforced online \cite{Belli2016}. However, this paradigm is failing, because of at least two reasons. 

First, as digital economies thrive, more and more concerns about consumer and citizen harms are native to the Internet and the information diffusion infrastructure that comes with it. Legal and ethical issues around consumer manipulation such as dark patterns \cite{DBLP:journals/corr/abs-1907-07032} centre around digital interfaces and user experience. Particularly in the case of this example, consumer organizations and national authorities have become increasingly interested in how existing rules on consumer manipulation apply to the context of using interface design choices to encourage commercial transactions. Yet one of the most important aspects of digital enforcement has been lagging behind in terms of its translation from legacy regulation to digital market regulation: the concept of sovereignty \cite{Eggenschwiler2017}. Legal enforcement is based on the idea that a sovereign state (or supranational organization deriving its legitimacy from national sovereignty) has the authority and legitimacy to set out its own rules and frameworks for creating sanctions when these rules are violated. The traditionally geographical dimension of sovereignty ensures the effectiveness for legal enforcement to a wide extent \cite{Kennedy2020}. Even in the physical space, due to the limitations of sovereignty, whenever legal activities develop a cross-border nature, hurdles around legal divergence and legitimate power arise. Looking outside its own borders has been difficult for the sovereign state. This is why the 'uploading' of physical borders into cyberspace has generally been a frustrating and frustrated process. Frustrating because even as recently as 2022, in order to pursue investigations about online violations of the law, national authorities need to navigate a complex web of procedural rules and rely on informal cross-border cooperation networks. At the same time, the establishment of the sovereign state in cyberspace has been frustrated by commercial and non-commercial actors that developed norms and practices with the goal of pushing against the division of the Internet space into national (or supranational) plots \cite{Belli2016}. 

Second, legal enforcement and scale on digital markets is a complex issue that might need the rethinking of what regulatory frameworks aim to achieve. With digital enforcement offering the possibility to make the invisible visible, societies around the world need to think long and hard about the goal of legal rules. Let us briefly reflect on an example from the physical world. 
For instance, towards the end of 2021, Dutch Prime Minister Mark Rutte declared on national television that most categories of fireworks were prohibited. This did not stop a considerable number of citizens from using fireworks, including the prohibited categories. If the police was expected to enforce the rules prohibiting the use of fireworks, it would have had the gargantuan task of starting investigations against tens if not hundreds of thousands of individuals, for the purpose of identifying those who broke the law and applying the relevant sanctions. However, in these situations, due to various reasons ranging from resource availability to public policy considerations, the police might end up enforcing the law only in a limited number of cases, while most of the infringements go unpunished. In civil matters, complaints about difficulties in accessing remedies such as obtaining a refund from a company that breached a consumer contract do not always reach the authorities tasked with the enforcement of consumer protection, or the courts supposed to impart justice in the case of an imbalanced transaction. This is due, once again, to a wide array of factors, including consumer behaviour or transaction costs. Injustices occurring in society often go generally unnoticed. However, technology can change that. The automation of legal enforcement could help in unveiling injustice at scale for the first time in human history, raising questions about the impact this would have on perception of justice as a social need, as well as the goal of drafting rules about human conduct and the sanctions that arise for their violations. 

So if the offline to online legal enforcement paradigm is failing, what exactly can replace it? To explore an answer to this question, in what follows we take a closer look at whether and to what extent digital enforcement can be automated in the context of the case studies presented above. 

\subsection{From enforcement goals to computational tasks}

Section~\ref{sec:scenarios} discussed two harms arising out of content monetization on social media, particularly affecting children as a vulnerable category of social media users. The goal of this section is to break down digital enforcement in a very practical discussion, and reflect on how to translate enforcement goals into computational tasks and identify what data would be needed for the completion of these tasks. The answer to this question has a two-fold structure. 

First, we tackle the more general aspects of digital enforcement, such as defining the actors who participate in the monetization ecosystem (particularly the vulnerable category we focus on in this paper) and their interaction: 

\begin{enumerate}
    \item \textit{Identifying under-aged users in order to restrict access to harmful content or apply stronger protections to their roles as creators.}
    
As a computational task, identifying child accounts in the current social media can be done in at least two ways: through self-disclosure or through inference. Self-disclosure is relevant as a computational parameter only if the information disclosed (e.g. date of birth) is truthful. Yet in some jurisdictions such as Germany, Facebook's real name policy as a requirement for the submission of truthful information was deemed illegal \cite{kastrenakes_2018}. Lying about identifying information such as name or age is particularly commonplace among children, since platforms sometimes restrict the creation of an account to persons over 13 (e.g. TikTok). 
This leaves age inference through content and interaction as a more reliable approach, which has been shown feasible~\cite{Zhang_Hu_Zhang_Liu_2021, farnadi2018profiling}. 
The data necessary for the completion of this task includes posted content (in various formats depending on the platform, thus including videos, pictures and text) and on-platform interactions with other users. 
Since scientists showed on limited data that demographic characteristics can be inferred, the platforms can/are probably doing already this for ad placement and other business-related objectives. 
Sharing this inferred data with authorities would provide transparency and improve the accountability of the platforms, in addition to contributing to digital law enforcement. 
Of course, concerns about governmental overreach in some cultures and regulatory environments will generate debates and search for new solutions. 


    \item \textit{Differentiating between commercial and non-commercial users to allocate legal obligations.}

Some platforms already log this user classification as metadata in their APIs, making it easy to identify commercial accounts. The DSA imposes additional requirements for the traceability of traders, such as a duty incumbent on platforms to obtain information about the legal persons they host who offer goods for sale. 
However, in the content creator economy, difficulties arise when determining for instance who is an influencer, and who engaged in these activities for monetization purposes and may thus qualify as a trader. 
TikTok has offered its creators the option to label their accounts as creator accounts. 
However, this is purely voluntary, and the development of algorithmic gossip~\cite{bishop2019alggossip} leads to concerns among influencers that using such categories can attract shadowbanning, making it a less than attractive choice. 
Similar to age inference, monetization inference can be based on contextual inference. 
Platforms could thus give access to two categories of data: self-reported classifications from verified traders (including companies and self-employed individuals), and inferred classifications based on its insights on account content. 
These labels can also help generate lists of users who fulfill certain criteria \cite{2021}, such as brands or influencers with a level of following determined relevant by enforcement authorities.

    \item \textit{Identifying connections at scale between various categories of users to map and visualise the ecosystem in which they operate.}

Digital enforcement needs reflect not only the possibility to retrieve data on individual users, but also to understand how these users interact in networks. 
While individual influencers who promore goods or content for child audiences may not individually raise flags related to online activities that should be subject to digital law enforcement, groups of influencers hired by the same company may have collectively a significant activity or audience.
Identifying such networks of influencers is not a trivial taks, and somewhat akin with coordinated influence operations~\cite{nimmo2020ira,Wai2021Multiplatform}.


\item \emph{Identifying user geographic data.}

In real-world enforcement, domicile is an essential concept to determine the applicable law to legal and natural persons. Just like age, geographic data can be voluntarily shared or inferred. Services offered by data intermediaries such as Heepsy can help delineate the pitfalls of inference, generally based on multiple data points. The language of posts is often not enough to profile geographical provenance or location, and data intermediaries turn to the language of comments, or the description of accounts. Especially in the case of the latter, flags are taken as proxies for geographical location, whereas flags (as emojis) can be used to communicate a different message (e.g. countries visited), or can be misused under the assumption that recommender systems can be gamed. However, as in previous cases of user demographic inferences, user location has been inferred even with the imperfect and limited data that scientists can collect from social media platforms~\cite{li2012home-location,ryoo2014-inferring-twitter-location, 10.1145/2505515.2505544}.  Platforms, however, have more data---from IP addresses, for example---that they use for ad placement, security verifications, etc.

\end{enumerate} 

Second, we identify enforcement goals that are sector-specific as well as dependent on applicable national and European legal frameworks. This is not a comprehensive or systematic enumeration, but rather an exemplification of the difficulties that may arise in automating enforcement. In this sense, we refer to a few observations based on the two scenarios discussed in Section~\ref{sec:scenarios}:

\begin{enumerate}
    \item \textit{Identifying how platforms comply with the obligation of dealing with a minor's right to delete own personal data.}

As indicated above, French law now mandates that platforms must 'facilitate the implementation, by minors, of the right to the
deletion of personal data [...] and to inform them, in clear and precise terms, easily understandable by them, of the modalities of implementation of this right'. In addition to parental consent tools, platforms must now also develop methods through which under-aged users from the French cyberspace ought to be able to manage their own consent, so as to delete posts made about them by a parental guardian who manages their economic activity as influencers. Content managed through such consent management infrastructures can be labelled as deleted by user. 

\item \textit{Retrieving user reports of audiovisual content that would undermine the dignity or moral or physical integrity of minors.}

French law also specifically integrated this platform obligation in its labour code. While some platforms allow users to report user accounts under 13 (e.g. TikTok), the reporting labels generally made available across platforms do not reflect a lot of granularity in terms of allowing users to flag content considered to be harmful for the dignity, moral or physical integrity of minors. Currently, reported content and content labels are not retrievable from platform APIs. Making this data available can help enforcement authorities determine compliance with the platform's obligation of facilitating such reports. 

    
\item \textit{Identifying non-disclosed advertising by under-aged influencers or targeting under-aged audiences.}

On social media, commercial and non-commercial content is flowing in user feeds with little to no consideration for the need to separate advertising from non-advertising messaging. Some platforms provide affordances such as 'paid partnership with' (e.g. Instagram), retrievable as metadata in the Instagram API. European consumer protection legislation and guidelines are universally converging on the acknowledgement that direct and indirect revenue ought to be disclosed. Apart from sharing voluntarily labels, it is not clear if platforms currently undertake internal measurements of predicted native advertising. Literature has recently proposed various models (supervised and semi-supervised) for the prediction of advertising in non-disclosed content. Such models can be used on data samples collected from the platform, with the caveat that collecting such data is always dependent on having a clear overview of the number of accounts over a specific size to be determined by authorities (e.g. top 1000 accounts with more than 1 million followers).

\end{enumerate}

\section{Discussion: The Paradoxes of Data Access}

So far in this paper, we looked at how digital enforcement can be achieved in two particular scenarios, providing data context for the computational tasks that are necessary for national authorities to verify legal compliance. This is not how enforcement generally takes place at the moment, although some jurisdictions are investing heavily in the development of data and behavioural units that can undertake forensic digital activities for the gathering of evidence of legal violations. However, if scaling enforcement is a desirable regulatory objective, and it can lead to a safer cyberspace, it is essential to consider how to properly scale it in a way that is consistent with procedural frameworks, on the one hand, and public interest, on the other. 

Chapter 4 of the DSA Proposal elaborates on the creation and activities of new institutions tasked with the enforcement of the obligations covered by the instrument. Article 31 in particular elaborates on obligations regarding data access applicable to very large platforms, defined in Article 25 as 'online platforms which provide their services to a number of average monthly active recipients of the service in the Union equal to or higher than 45 million, calculated in accordance with the methodology set out in the delegated acts referred to in paragraph 3'. Article 31(1) sets out an obligation for very large platforms to provide the proposed national Digital Services Coordinators or the Commission 'upon their reasoned request and within a reasonable period, specified in the request, access to data that are necessary to monitor and assess compliance with this Regulation', which will be used only for these enforcement purposes. In addition, Article 31(3) further mentions that very large online platforms 'shall provide access to data pursuant to paragraphs 1 and 2 through online databases or application programming interfaces (APIs), as appropriate'. APIs are also referred to as a desirable voluntary industry standard in Article 34, elaborated upon in Recital 66 of the Preamble, stating that 'industry can help develop standardised means to comply with this Regulation, such as allowing the submission of notices, including through application programming interfaces [...]. The standards could distinguish between different types of illegal content or different types of intermediary services, as appropriate.' This overview exhausts the references to data access through platform APIs in the DSA Proposal, proving that while some thought was given to the technological enforcement of the DSA, a coherent standard for data access is not yet in sight. Article 31(5) indicates that the Commission will adopt delegated acts to elaborate on 'the technical conditions under which very large online platforms are to share data pursuant to paragraphs 1 and 2 and the purposes for which the data may be used'. 

This paper aims to be a starting point for a discussion of what the delegated acts can achieve in terms of standardizing data access for the purpose of legal compliance. It must be noted that although Article 31 also deals with data access by researchers \cite{Leerssen}, such activities do not count as legal enforcement . We propose the concept of a legal compliance API which can be used for the enforcement of the DSA, and this section elaborates on potential characteristics, benefits and limitations of such a concept.  

\begin{enumerate}
    \item \textit{The legal compliance API should be unique and available to competent regulatory authorities across sectors}
    
As we have seen in Section~\ref{sec:scenarios}, digital enforcement is now a need in any field of European and national regulation that deals with digital markets. Scenario 1 showed how French authorities tasked with the enforcement of child labour protections will need to undertake investigations to apply recent rules relating to child influencers. Although this law has only been adopted in one Member State so far, it is only a matter of time until other jurisdictions and authorities will follow suit. Similarly, scenario 2 discussed the enforcement of legal standards relating to surreptitious advertising by underaged influencers or to underaged audiences. In this scenario, consumer and media authorities at national level compete for the enforcement of instruments such as the UCPD. 
In addition, it must be noted that in the case of consumer protection, additional digital enforcement powers are relevant, as per the Consumer Protection Cooperation Regulation 2017, which grants consumer authorities generous investigation and enforcement powers relating to data access or website takedowns, just to mention a few examples, as well as the powers to ask for data from any other relevant authority. What this shows is a proliferation of interest in data access across a wide array of national regulators. To account for this in a digital enforcement infrastructure, a legal compliance API would have to be unique and available to any national authority with legitimate powers of investigation and enforcement. To reiterate, the DSA Proposal makes it clear, in its definition of illegal content, that sectoral regulation is a source of defining this illegality, making the DSA a bridge between the procedural frameworks it sets up, and additional national and European legislation relevant for online content.

\item \textit{The legal compliance API should allow competent regulatory authorities to define relevant data for the purpose of data access.}
    
Voluntary APIs provided by platforms currently have serious limitations which trickle down into discussions relating transaction costs for enforcement, such as the computational power to use data for the investigation of legal violations. For instance, social media platforms do not include in their API documentations any query parameter that can allow for the retrieval of data on users ranked according to factors such as size of the follower base. Through a legal compliance API independent of any available platform APIs, national authorities could propose API parameters that may be already retrievable based on data labels not disclosed in existing specifications (e.g. reported content). Apart from undisclosed data labels which would entail minimal transaction costs for platforms, authorities could also set out parameters for queries on the basis of content that platforms process. An example here would be age labels for accounts based on age prediction models already employed by platforms. This latter option can allow for authorities to outsource computational costs to platforms, who are best positioned to run such models internally. However, this also raises questions relating the vetting of the models, and platform liability standards if age prediction algorithms are so deficient that they fail to enact the protection of minors as defined by law. Algorithmic architectures and systemic risks are a meta category of legal compliance which needs additional comprehensive analysis. The scenarios covered in this paper, and the legal enforcement discussions raised around them reflect more targeted interventions, where platforms may have some incentives to collaborate with authorities as intermediaries.

\item \textit{The legal compliance API should be subject to auditing and additional compliance checks. }

As noted above, even if data can be collected via API, they are not guaranteed to be accurate because platforms are in a unique position to have full oversight over own data, whereas any external entity, including regulatory authorities, does not. Recent scandals relating to inaccurate data offered by Facebook to researchers only amplify the worries that when asked for data access which may compromise them, platforms may have serious incentives to skew data \cite{fbsso}. Similarly, research has shown that APIs available for research purposes where samples of data are shared (e.g. Twitter's 1\% free API), can offer skewed representations of the whole data available \cite{pfeffer2018tampering}. This is a fundamental issue for the scalability of digital compliance, yet there are at least a few ways in which some of the concerns surrounding data accuracy can be remedied. Inspiration is available from fields where investigation and enforcement powers have considerably increased in the past years, such as consumer protection. In the EU, consumer authorities undertake what is called 'mystery shopping' to verify legal compliance. In other words, they randomly buy goods or services from traders under false pretenses, to verify how traders comply with their legal obligations. A similar approach can be taken through a triangulation of the various APIs made available by platforms, retrieved under false pretense in a lawful manner in order to test data accuracy. Moreover, accuracy concerns can also be addressed in procedural frameworks. For instance, placing the burden of proof on platforms in showing their compliance or non-compliance with legal obligations can shift proof to the party which has most data.

\item \textit{The legal compliance API should be governed by responsible rules of public administration accountability.}

Digital enforcement entails that not only companies are able to track online activities, but also public authorities. In various jurisdictions, this has caused a considerable backlash because of surveillance concerns, as often legal procedural frameworks around Internet investigations by public authorities are not only simply missing, but existing regulation does not account for any degree of transparency relating when and how authorities can exercise surveillance powers, and how these powers are kept in check. This is evidently a great concern that needs to be tackled in any case, given the exponential interest in digital enforcement which is gradually turning into a public need that online markets cannot do without. For a legal compliance API to be a viable solution, additional standards are necessary to improve the transparency of how public institutions handle data, and redress mechanisms for when lawful powers are exercised in unlawful ways. In some fields of regulation, such as data or consumer protection, this is an ongoing debate focused very much on procedural and administrative law issues, such as the admissibility of evidence collected online, as well as informal coordination practices such as data sharing standards applicable to regulatory cooperation networks. 

\end{enumerate}

As already mentioned, for many pieces of information, platforms already have much of the inferred data required by a legal API. 
However, in some other cases the platforms are overwhelmed by very significant issues (such as identifying state-run information operations and disinformation campaigns) where they need to detect networks of accounts which aim to manipulate segments of the population. It is difficult to make the case to put more computational burden on platforms to, for example, also identify networks of accounts that collectively violate laws in various parts of the world. At the same time, public authorities have the powers to pursue investigations relating to violations of the law; however, authorities enforcing consumer or social protection are not native to the space of digital evidence. Adding to this problems of resource availability, organizational innovation or information literacy considerably increases this burden. However, scientists all over the globe can help monitor and identify legal compliance issues with easier access to data, which the DSA is committing to support. Additionally, in this respect, the recent legislative proposal~\cite{PATA2021} in the US to give unlimited social media access to research projects selected for funding by the US National Science Foundation is a very significant step towards allowing scientists to help platforms monitor themselves for social good.

The legal compliance API as laid out in this paper is a mere concept that needs considerable additional analysis and reflection. As we have seen, it is by no means a complete solution. Yet in the current state of affairs, authorities have to deal with a paralyzing information asymmetry which affects their ability to pierce the veil of platform data. A legal compliance API offers significant benefits because it can offer opportunities for legal enforcement which currently do not exist. Further research can build on e.g. other sectors, technical standards or potential specifications for such an API, as well as behavioural questions relating how such enforcement transparency could distort user and platform conduct.

\bibliographystyle{ACM-Reference-Format}
\bibliography{main.bib}


\begin{thebibliography}{45}


\ifx \showCODEN    \undefined \def \showCODEN     #1{\unskip}     \fi
\ifx \showDOI      \undefined \def \showDOI       #1{#1}\fi
\ifx \showISBNx    \undefined \def \showISBNx     #1{\unskip}     \fi
\ifx \showISBNxiii \undefined \def \showISBNxiii  #1{\unskip}     \fi
\ifx \showISSN     \undefined \def \showISSN      #1{\unskip}     \fi
\ifx \showLCCN     \undefined \def \showLCCN      #1{\unskip}     \fi
\ifx \shownote     \undefined \def \shownote      #1{#1}          \fi
\ifx \showarticletitle \undefined \def \showarticletitle #1{#1}   \fi
\ifx \showURL      \undefined \def \showURL       {\relax}        \fi
\providecommand\bibfield[2]{#2}
\providecommand\bibinfo[2]{#2}
\providecommand\natexlab[1]{#1}
\providecommand\showeprint[2][]{arXiv:#2}

\bibitem[\protect\citeauthoryear{??}{noa}{[n.\,d.]a}]%
        {noauthor_digital_nodate-1}
 \bibinfo{year}{[n.\,d.]}\natexlab{a}.
\newblock \bibinfo{title}{The {Digital} {Millennium} {Copyright} {Act}
  {\textbar} {U}.{S}. {Copyright} {Office}}.
\newblock
\newblock
\urldef\tempurl%
\url{https://www.copyright.gov/dmca/}
\showURL{%
\tempurl}


\bibitem[\protect\citeauthoryear{??}{noa}{[n.\,d.]b}]%
        {noauthor_digital_nodate}
 \bibinfo{year}{[n.\,d.]}\natexlab{b}.
\newblock \bibinfo{title}{The {Digital} {Services} {Act} package {\textbar}
  {Shaping} {Europe}’s digital future}.
\newblock
\newblock
\urldef\tempurl%
\url{https://digital-strategy.ec.europa.eu/en/policies/digital-services-act-package}
\showURL{%
\tempurl}


\bibitem[\protect\citeauthoryear{??}{noa}{[n.\,d.]c}]%
        {noauthor_eur-lex_nodate-1}
 \bibinfo{year}{[n.\,d.]}\natexlab{c}.
\newblock \bibinfo{title}{{EUR}-{Lex} - {32016R0679} - {EN} - {EUR}-{Lex}}.
\newblock
\newblock
\urldef\tempurl%
\url{https://eur-lex.europa.eu/eli/reg/2016/679/oj}
\showURL{%
\tempurl}


\bibitem[\protect\citeauthoryear{??}{noa}{[n.\,d.]d}]%
        {noauthor_eur-lex_nodate}
 \bibinfo{year}{[n.\,d.]}\natexlab{d}.
\newblock \bibinfo{title}{{EUR}-{Lex} - {52021PC0206} - {EN} - {EUR}-{Lex}}.
\newblock
\newblock
\urldef\tempurl%
\url{https://eur-lex.europa.eu/legal-content/EN/TXT/?uri=CELEX%3A52021PC0206}
\showURL{%
\tempurl}


\bibitem[\protect\citeauthoryear{??}{noa}{[n.\,d.]e}]%
        {noauthor_eur-lex_nodate-2}
 \bibinfo{year}{[n.\,d.]}\natexlab{e}.
\newblock \bibinfo{title}{{EUR}-{Lex} - {62020CJ0371} - {EN} - {EUR}-{Lex}}.
\newblock
\newblock
\urldef\tempurl%
\url{https://eur-lex.europa.eu/legal-content/EN/TXT/?uri=CELEX%3A62020CJ0371}
\showURL{%
\tempurl}


\bibitem[\protect\citeauthoryear{??}{noa}{[n.\,d.]f}]%
        {noauthor_monetisation_nodate}
 \bibinfo{year}{[n.\,d.]}\natexlab{f}.
\newblock \bibinfo{title}{Monetisation tools application {\textbar} {Meta} for
  {Creators}}.
\newblock
\newblock
\urldef\tempurl%
\url{https://www.facebook.com/creators/tools/mta}
\showURL{%
\tempurl}


\bibitem[\protect\citeauthoryear{??}{PAT}{[n.\,d.]}]%
        {PATA2021}
 \bibinfo{year}{[n.\,d.]}\natexlab{}.
\newblock \bibinfo{title}{The {P}latform {A}ccountability and {T}ransparency
  {A}ct}.
\newblock
\newblock
\urldef\tempurl%
\url{https://www.coons.senate.gov/imo/media/doc/text_pata_117.pdf}
\showURL{%
\tempurl}


\bibitem[\protect\citeauthoryear{??}{noa}{[n.\,d.]g}]%
        {noauthor_press_nodate}
 \bibinfo{year}{[n.\,d.]}\natexlab{g}.
\newblock \bibinfo{title}{Press corner}.
\newblock
\newblock
\urldef\tempurl%
\url{https://ec.europa.eu/commission/presscorner/home/en}
\showURL{%
\tempurl}


\bibitem[\protect\citeauthoryear{Belli and Venturini}{Belli and
  Venturini}{2016}]%
        {Belli2016}
\bibfield{author}{\bibinfo{person}{Luca Belli} {and} \bibinfo{person}{Jamila
  Venturini}.} \bibinfo{year}{2016}\natexlab{}.
\newblock \showarticletitle{Private ordering and the rise of terms of service
  as cyber-regulation}.
\newblock \bibinfo{journal}{\emph{Internet Policy Review}} \bibinfo{volume}{5},
  \bibinfo{number}{4} (\bibinfo{date}{Dec.} \bibinfo{year}{2016}).
\newblock
\urldef\tempurl%
\url{https://doi.org/10.14763/2016.4.441}
\showDOI{\tempurl}


\bibitem[\protect\citeauthoryear{Bertaglia, Grigoriu, Dumontier, and van
  Dijck}{Bertaglia et~al\mbox{.}}{2021}]%
        {bertaglia-etal-2021-abusive}
\bibfield{author}{\bibinfo{person}{Thales Bertaglia}, \bibinfo{person}{Andreea
  Grigoriu}, \bibinfo{person}{Michel Dumontier}, {and} \bibinfo{person}{Gijs
  van Dijck}.} \bibinfo{year}{2021}\natexlab{}.
\newblock \showarticletitle{Abusive Language on Social Media Through the Legal
  Looking Glass}. In \bibinfo{booktitle}{\emph{Proceedings of the 5th Workshop
  on Online Abuse and Harms (WOAH 2021)}}. \bibinfo{publisher}{Association for
  Computational Linguistics}, \bibinfo{address}{Online},
  \bibinfo{pages}{191--200}.
\newblock
\urldef\tempurl%
\url{https://doi.org/10.18653/v1/2021.woah-1.20}
\showDOI{\tempurl}


\bibitem[\protect\citeauthoryear{Bishop}{Bishop}{2019}]%
        {bishop2019alggossip}
\bibfield{author}{\bibinfo{person}{Sophie Bishop}.}
  \bibinfo{year}{2019}\natexlab{}.
\newblock \showarticletitle{Managing visibility on YouTube through algorithmic
  gossip}.
\newblock \bibinfo{journal}{\emph{New Media \& Society}} \bibinfo{volume}{21},
  \bibinfo{number}{11-12} (\bibinfo{year}{2019}), \bibinfo{pages}{2589--2606}.
\newblock
\urldef\tempurl%
\url{https://doi.org/10.1177/1461444819854731}
\showDOI{\tempurl}
\showeprint{https://doi.org/10.1177/1461444819854731}


\bibitem[\protect\citeauthoryear{Campbell}{Campbell}{2016}]%
        {campbell2016}
\bibfield{author}{\bibinfo{person}{Angela~J. Campbell}.}
  \bibinfo{year}{2016}\natexlab{}.
\newblock \showarticletitle{Rethinking Children's Advertising Policies for the
  Digital Age}.
\newblock \bibinfo{journal}{\emph{Loyola Consumer Law Review}}
  \bibinfo{volume}{29}, \bibinfo{number}{1} (\bibinfo{year}{2016}),
  \bibinfo{pages}{1--54}.
\newblock


\bibitem[\protect\citeauthoryear{CAUFFMAN and GOANTA}{CAUFFMAN and
  GOANTA}{2021}]%
        {cauffman_goanta_2021}
\bibfield{author}{\bibinfo{person}{Caroline CAUFFMAN} {and}
  \bibinfo{person}{Catalina GOANTA}.} \bibinfo{year}{2021}\natexlab{}.
\newblock \showarticletitle{A New Order: The Digital Services Act and Consumer
  Protection}.
\newblock \bibinfo{journal}{\emph{European Journal of Risk Regulation}}
  \bibinfo{volume}{12}, \bibinfo{number}{4} (\bibinfo{year}{2021}),
  \bibinfo{pages}{758–774}.
\newblock
\urldef\tempurl%
\url{https://doi.org/10.1017/err.2021.8}
\showDOI{\tempurl}


\bibitem[\protect\citeauthoryear{Duivenvoorde}{Duivenvoorde}{2022}]%
        {Duivenvoorde2022}
\bibfield{author}{\bibinfo{person}{Bram Duivenvoorde}.}
  \bibinfo{year}{2022}\natexlab{}.
\newblock \showarticletitle{The Liability of Online Marketplaces under the
  Unfair Commercial Practices Directive, the E-commerce Directive and the
  Digital Services Act}.
\newblock \bibinfo{journal}{\emph{Journal of European Consumer and Market Law}}
  \bibinfo{volume}{11}, \bibinfo{number}{2} (\bibinfo{year}{2022}),
  \bibinfo{pages}{43--52}.
\newblock
\urldef\tempurl%
\url{https://kluwerlawonline.com/journalarticle/Journal+of+European+Consumer+and+Market+Law/11.2/EuCML2022009}
\showURL{%
\tempurl}


\bibitem[\protect\citeauthoryear{Eggenschwiler}{Eggenschwiler}{2017}]%
        {Eggenschwiler2017}
\bibfield{author}{\bibinfo{person}{Jacqueline Eggenschwiler}.}
  \bibinfo{year}{2017}\natexlab{}.
\newblock \showarticletitle{Accountability challenges confronting cyberspace
  governance}.
\newblock \bibinfo{journal}{\emph{Internet Policy Review}} \bibinfo{volume}{6},
  \bibinfo{number}{3} (\bibinfo{date}{Sept.} \bibinfo{year}{2017}).
\newblock
\urldef\tempurl%
\url{https://doi.org/10.14763/2017.3.712}
\showDOI{\tempurl}


\bibitem[\protect\citeauthoryear{Elwood, Gasparin, and Rozza}{Elwood
  et~al\mbox{.}}{2021}]%
        {2021}
\bibfield{author}{\bibinfo{person}{Adam Elwood}, \bibinfo{person}{Alberto
  Gasparin}, {and} \bibinfo{person}{Alessandro Rozza}.}
  \bibinfo{year}{2021}\natexlab{}.
\newblock \showarticletitle{Ranking Micro-Influencers: a Novel Multi-Task
  Learning and Interpretable Framework}.
\newblock \bibinfo{journal}{\emph{2021 IEEE International Symposium on
  Multimedia (ISM)}} (\bibinfo{date}{Nov} \bibinfo{year}{2021}).
\newblock
\urldef\tempurl%
\url{https://doi.org/10.1109/ism52913.2021.00030}
\showDOI{\tempurl}


\bibitem[\protect\citeauthoryear{{European Commission}}{{European
  Commission}}{2020}]%
        {euwhitepaper2020}
\bibfield{author}{\bibinfo{person}{{European Commission}}.}
  \bibinfo{year}{2020}\natexlab{}.
\newblock \bibinfo{booktitle}{\emph{White Paper on Artificial Intelligence: a
  European approach to excellence and trust}}.
\newblock \bibinfo{type}{White Paper} COM(2020) 65 final.
  \bibinfo{institution}{European Commission}, \bibinfo{address}{Brussels}.
\newblock
\urldef\tempurl%
\url{https://ec.europa.eu/info/files/white-paper-artificial-intelligence-european-approach-excellence-and-trust_en}
\showURL{%
\tempurl}


\bibitem[\protect\citeauthoryear{Farnadi, Tang, De~Cock, and Moens}{Farnadi
  et~al\mbox{.}}{2018}]%
        {farnadi2018profiling}
\bibfield{author}{\bibinfo{person}{Golnoosh Farnadi}, \bibinfo{person}{Jie
  Tang}, \bibinfo{person}{Martine De~Cock}, {and}
  \bibinfo{person}{Marie-Francine Moens}.} \bibinfo{year}{2018}\natexlab{}.
\newblock \showarticletitle{User Profiling through Deep Multimodal Fusion}. In
  \bibinfo{booktitle}{\emph{Proceedings of the Eleventh ACM International
  Conference on Web Search and Data Mining}} (Marina Del Rey, CA, USA)
  \emph{(\bibinfo{series}{WSDM '18})}. \bibinfo{publisher}{Association for
  Computing Machinery}, \bibinfo{address}{New York, NY, USA},
  \bibinfo{pages}{171–179}.
\newblock
\showISBNx{9781450355810}
\urldef\tempurl%
\url{https://doi.org/10.1145/3159652.3159691}
\showDOI{\tempurl}


\bibitem[\protect\citeauthoryear{Goanta, Louisse, and Ortolani}{Goanta
  et~al\mbox{.}}{2021}]%
        {goantaOrtolani2021}
\bibfield{author}{\bibinfo{person}{Catalina Goanta}, \bibinfo{person}{Marije
  Louisse}, {and} \bibinfo{person}{Pietro Ortolani}.}
  \bibinfo{year}{2021}\natexlab{}.
\newblock \bibinfo{booktitle}{\emph{Marketing Communications and the Digital
  Single Market}}.
\newblock \bibinfo{publisher}{Oxford University Press},
  \bibinfo{address}{United Kingdom}, \bibinfo{pages}{287--298}.
\newblock
\showISBNx{9780192856395}


\bibitem[\protect\citeauthoryear{Gorwa}{Gorwa}{2019}]%
        {Gorwa2019}
\bibfield{author}{\bibinfo{person}{Robert Gorwa}.}
  \bibinfo{year}{2019}\natexlab{}.
\newblock \showarticletitle{The platform governance triangle: conceptualising
  the informal regulation of online content}.
\newblock \bibinfo{journal}{\emph{Internet Policy Review}} \bibinfo{volume}{8},
  \bibinfo{number}{2} (\bibinfo{date}{June} \bibinfo{year}{2019}).
\newblock
\urldef\tempurl%
\url{https://doi.org/10.14763/2019.2.1407}
\showDOI{\tempurl}


\bibitem[\protect\citeauthoryear{Hills}{Hills}{[n.\,d.]}]%
        {hills_information_nodate}
\bibfield{author}{\bibinfo{person}{Filippo~Menczer Hills, Thomas}.}
  \bibinfo{year}{[n.\,d.]}\natexlab{}.
\newblock \bibinfo{title}{Information {Overload} {Helps} {Fake} {News}
  {Spread}, and {Social} {Media} {Knows} {It}}.
\newblock
\newblock
\urldef\tempurl%
\url{https://doi.org/10.1038/scientificamerican1220-54}
\showDOI{\tempurl}


\bibitem[\protect\citeauthoryear{James}{James}{2017}]%
        {james2017}
\bibfield{author}{\bibinfo{person}{Tisha James}.}
  \bibinfo{year}{2017}\natexlab{}.
\newblock \showarticletitle{The Real Sponsors of Social Media: How Internet
  Influencers Are Escaping FTC Disclosure Laws}.
\newblock \bibinfo{journal}{\emph{Ohio State Business Law Journal}}
  \bibinfo{volume}{11}, \bibinfo{number}{1} (\bibinfo{year}{2017}),
  \bibinfo{pages}{61--86}.
\newblock


\bibitem[\protect\citeauthoryear{Kastrenakes}{Kastrenakes}{2018}]%
        {kastrenakes_2018}
\bibfield{author}{\bibinfo{person}{Jacob Kastrenakes}.}
  \bibinfo{year}{2018}\natexlab{}.
\newblock \bibinfo{title}{German court says Facebook's real name policy is
  illegal}.
\newblock
\newblock
\urldef\tempurl%
\url{https://www.theverge.com/2018/2/12/17005746/facebook-real-name-policy-illegal-german-court-rules}
\showURL{%
\tempurl}


\bibitem[\protect\citeauthoryear{Kennedy and Warren}{Kennedy and
  Warren}{2020}]%
        {Kennedy2020}
\bibfield{author}{\bibinfo{person}{Sally Kennedy} {and} \bibinfo{person}{Ian
  Warren}.} \bibinfo{year}{2020}\natexlab{}.
\newblock \showarticletitle{The legal geographies of extradition and sovereign
  power}.
\newblock \bibinfo{journal}{\emph{Internet Policy Review}}
  (\bibinfo{year}{2020}).
\newblock
\urldef\tempurl%
\url{https://doi.org/10.14763/2020.3.1496}
\showDOI{\tempurl}


\bibitem[\protect\citeauthoryear{Leerssen}{Leerssen}{[n.\,d.]}]%
        {Leerssen}
\bibfield{author}{\bibinfo{person}{Paddy Leerssen}.}
  \bibinfo{year}{[n.\,d.]}\natexlab{}.
\newblock \bibinfo{title}{Platform research access in {Article} 31 of the
  {Digital} {Services} {Act}}.
\newblock
\newblock
\urldef\tempurl%
\url{https://verfassungsblog.de/power-dsa-dma-14/}
\showURL{%
\tempurl}


\bibitem[\protect\citeauthoryear{Li, Wang, Deng, Wang, and Chang}{Li
  et~al\mbox{.}}{2012}]%
        {li2012home-location}
\bibfield{author}{\bibinfo{person}{Rui Li}, \bibinfo{person}{Shengjie Wang},
  \bibinfo{person}{Hongbo Deng}, \bibinfo{person}{Rui Wang}, {and}
  \bibinfo{person}{Kevin Chen-Chuan Chang}.} \bibinfo{year}{2012}\natexlab{}.
\newblock \showarticletitle{Towards Social User Profiling: Unified and
  Discriminative Influence Model for Inferring Home Locations}. In
  \bibinfo{booktitle}{\emph{Proceedings of the 18th ACM SIGKDD International
  Conference on Knowledge Discovery and Data Mining}} (Beijing, China)
  \emph{(\bibinfo{series}{KDD '12})}. \bibinfo{publisher}{Association for
  Computing Machinery}, \bibinfo{address}{New York, NY, USA},
  \bibinfo{pages}{1023–1031}.
\newblock
\showISBNx{9781450314626}
\urldef\tempurl%
\url{https://doi.org/10.1145/2339530.2339692}
\showDOI{\tempurl}


\bibitem[\protect\citeauthoryear{Mathur, Acar, Friedman, Lucherini, Mayer,
  Chetty, and Narayanan}{Mathur et~al\mbox{.}}{2019}]%
        {DBLP:journals/corr/abs-1907-07032}
\bibfield{author}{\bibinfo{person}{Arunesh Mathur}, \bibinfo{person}{Gunes
  Acar}, \bibinfo{person}{Michael Friedman}, \bibinfo{person}{Elena Lucherini},
  \bibinfo{person}{Jonathan~R. Mayer}, \bibinfo{person}{Marshini Chetty}, {and}
  \bibinfo{person}{Arvind Narayanan}.} \bibinfo{year}{2019}\natexlab{}.
\newblock \showarticletitle{Dark Patterns at Scale: Findings from a Crawl of
  11K Shopping Websites}.
\newblock \bibinfo{journal}{\emph{CoRR}}  \bibinfo{volume}{abs/1907.07032}
  (\bibinfo{year}{2019}).
\newblock
\showeprint[arXiv]{1907.07032}
\urldef\tempurl%
\url{http://arxiv.org/abs/1907.07032}
\showURL{%
\tempurl}


\bibitem[\protect\citeauthoryear{Mathur, Narayanan, and Chetty}{Mathur
  et~al\mbox{.}}{2018}]%
        {2018}
\bibfield{author}{\bibinfo{person}{Arunesh Mathur}, \bibinfo{person}{Arvind
  Narayanan}, {and} \bibinfo{person}{Marshini Chetty}.}
  \bibinfo{year}{2018}\natexlab{}.
\newblock \showarticletitle{Endorsements on Social Media}.
\newblock \bibinfo{journal}{\emph{Proceedings of the ACM on Human-Computer
  Interaction}} \bibinfo{volume}{2}, \bibinfo{number}{CSCW}
  (\bibinfo{date}{Nov} \bibinfo{year}{2018}), \bibinfo{pages}{1–26}.
\newblock
\showISSN{2573-0142}
\urldef\tempurl%
\url{https://doi.org/10.1145/3274388}
\showDOI{\tempurl}


\bibitem[\protect\citeauthoryear{McGee, Caverlee, and Cheng}{McGee
  et~al\mbox{.}}{2013}]%
        {10.1145/2505515.2505544}
\bibfield{author}{\bibinfo{person}{Jeffrey McGee}, \bibinfo{person}{James
  Caverlee}, {and} \bibinfo{person}{Zhiyuan Cheng}.}
  \bibinfo{year}{2013}\natexlab{}.
\newblock \showarticletitle{Location Prediction in Social Media Based on Tie
  Strength}. In \bibinfo{booktitle}{\emph{Proceedings of the 22nd ACM
  International Conference on Information \& Knowledge Management}} (San
  Francisco, California, USA) \emph{(\bibinfo{series}{CIKM '13})}.
  \bibinfo{publisher}{Association for Computing Machinery},
  \bibinfo{address}{New York, NY, USA}, \bibinfo{pages}{459–468}.
\newblock
\showISBNx{9781450322638}
\urldef\tempurl%
\url{https://doi.org/10.1145/2505515.2505544}
\showDOI{\tempurl}


\bibitem[\protect\citeauthoryear{NG, Horawalavithana, and Iamnitchi}{NG
  et~al\mbox{.}}{2021}]%
        {Wai2021Multiplatform}
\bibfield{author}{\bibinfo{person}{Kin~Wai NG}, \bibinfo{person}{Sameera
  Horawalavithana}, {and} \bibinfo{person}{Adriana Iamnitchi}.}
  \bibinfo{year}{2021}\natexlab{}.
\newblock \showarticletitle{{Multi-platform Information Operations: Twitter,
  Facebook and YouTube against the White Helmets}}. In
  \bibinfo{booktitle}{\emph{Proceedings of The Workshop Proceedings of the 14th
  International AAAI Conference on Web and Social Media (ICWSM)
  (SocialSens'21)}}. \bibinfo{address}{Atlanta, USA}.
\newblock


\bibitem[\protect\citeauthoryear{Nimmo, Francois, Eib, and Ronzaud}{Nimmo
  et~al\mbox{.}}{2020}]%
        {nimmo2020ira}
\bibfield{author}{\bibinfo{person}{Ben Nimmo}, \bibinfo{person}{Camille
  Francois}, \bibinfo{person}{C~Shawn Eib}, {and} \bibinfo{person}{Lea
  Ronzaud}.} \bibinfo{year}{2020}\natexlab{}.
\newblock \showarticletitle{IRA again: Unlucky thirteen}.
\newblock \bibinfo{journal}{\emph{Graphika, September}}  \bibinfo{volume}{1}
  (\bibinfo{year}{2020}).
\newblock


\bibitem[\protect\citeauthoryear{Park and Chung}{Park and Chung}{2012}]%
        {Park_2012}
\bibfield{author}{\bibinfo{person}{Jaimie~Y. Park} {and}
  \bibinfo{person}{Chin-Wan Chung}.} \bibinfo{year}{2012}\natexlab{}.
\newblock \showarticletitle{When Daily Deal Services Meet Twitter:
  Understanding Twitter as a Daily Deal Marketing Platform}. In
  \bibinfo{booktitle}{\emph{Proceedings of the 4th Annual ACM Web Science
  Conference}} (Evanston, Illinois) \emph{(\bibinfo{series}{WebSci '12})}.
  \bibinfo{publisher}{Association for Computing Machinery},
  \bibinfo{address}{New York, NY, USA}, \bibinfo{pages}{233–242}.
\newblock
\showISBNx{9781450312288}
\urldef\tempurl%
\url{https://doi.org/10.1145/2380718.2380748}
\showDOI{\tempurl}


\bibitem[\protect\citeauthoryear{Peukert, Husovec, Kretschmer, Mezei, and
  Quintais}{Peukert et~al\mbox{.}}{2022}]%
        {Peukert2022}
\bibfield{author}{\bibinfo{person}{Alexander Peukert}, \bibinfo{person}{Martin
  Husovec}, \bibinfo{person}{Martin Kretschmer}, \bibinfo{person}{P{\'{e}}ter
  Mezei}, {and} \bibinfo{person}{Jo{\~{a}}o~Pedro Quintais}.}
  \bibinfo{year}{2022}\natexlab{}.
\newblock \showarticletitle{European Copyright Society {\textendash} Comment on
  Copyright and the Digital Services Act Proposal}.
\newblock \bibinfo{journal}{\emph{{IIC} - International Review of Intellectual
  Property and Competition Law}} \bibinfo{volume}{53}, \bibinfo{number}{3}
  (\bibinfo{date}{March} \bibinfo{year}{2022}), \bibinfo{pages}{358--376}.
\newblock
\urldef\tempurl%
\url{https://doi.org/10.1007/s40319-022-01154-1}
\showDOI{\tempurl}


\bibitem[\protect\citeauthoryear{{Pfeffer, J\"urgen}, {Mayer, Katja}, and
  {Morstatter, Fred}}{{Pfeffer, J\"urgen} et~al\mbox{.}}{2018}]%
        {pfeffer2018tampering}
\bibfield{author}{\bibinfo{person}{{Pfeffer, J\"urgen}},
  \bibinfo{person}{{Mayer, Katja}}, {and} \bibinfo{person}{{Morstatter,
  Fred}}.} \bibinfo{year}{2018}\natexlab{}.
\newblock \showarticletitle{Tampering with Twitter\'{}s Sample API}.
\newblock \bibinfo{journal}{\emph{EPJ Data Sci.}} \bibinfo{volume}{7},
  \bibinfo{number}{1} (\bibinfo{year}{2018}), \bibinfo{pages}{50}.
\newblock
\urldef\tempurl%
\url{https://doi.org/10.1140/epjds/s13688-018-0178-0}
\showDOI{\tempurl}


\bibitem[\protect\citeauthoryear{Pottinger}{Pottinger}{2018}]%
        {pottinger2018}
\bibfield{author}{\bibinfo{person}{Nicole~E. Pottinger}.}
  \bibinfo{year}{2018}\natexlab{}.
\newblock \showarticletitle{Don't Forget to Subscribe: Regulation of Online
  Advertising Evaluated through YouTube's Monetization Problem}.
\newblock \bibinfo{journal}{\emph{Kentucky Law Journal}} \bibinfo{volume}{107},
  \bibinfo{number}{3} (\bibinfo{year}{2018}), \bibinfo{pages}{515--[viii]}.
\newblock


\bibitem[\protect\citeauthoryear{PURNHAGEN}{PURNHAGEN}{2017}]%
        {purnhagen_2017}
\bibfield{author}{\bibinfo{person}{Kai PURNHAGEN}.}
  \bibinfo{year}{2017}\natexlab{}.
\newblock \showarticletitle{More Reality in the CJEU’s Interpretation of the
  Average Consumer Benchmark – Also More Behavioural Science in Unfair
  Commercial Practices?}
\newblock \bibinfo{journal}{\emph{European Journal of Risk Regulation}}
  \bibinfo{volume}{8}, \bibinfo{number}{2} (\bibinfo{year}{2017}),
  \bibinfo{pages}{437–440}.
\newblock
\urldef\tempurl%
\url{https://doi.org/10.1017/err.2017.13}
\showDOI{\tempurl}


\bibitem[\protect\citeauthoryear{Reddy, Yu, Pappu, Sivaraman, Rezapour, and
  Jones}{Reddy et~al\mbox{.}}{2021}]%
        {reddy-etal-2021-detecting}
\bibfield{author}{\bibinfo{person}{Sravana Reddy}, \bibinfo{person}{Yongze Yu},
  \bibinfo{person}{Aasish Pappu}, \bibinfo{person}{Aswin Sivaraman},
  \bibinfo{person}{Rezvaneh Rezapour}, {and} \bibinfo{person}{Rosie Jones}.}
  \bibinfo{year}{2021}\natexlab{}.
\newblock \showarticletitle{Detecting Extraneous Content in Podcasts}. In
  \bibinfo{booktitle}{\emph{Proceedings of the 16th Conference of the European
  Chapter of the Association for Computational Linguistics: Main Volume}}.
  \bibinfo{publisher}{Association for Computational Linguistics},
  \bibinfo{address}{Online}, \bibinfo{pages}{1166--1173}.
\newblock
\urldef\tempurl%
\url{https://doi.org/10.18653/v1/2021.eacl-main.99}
\showDOI{\tempurl}


\bibitem[\protect\citeauthoryear{Rieder and Hofmann}{Rieder and
  Hofmann}{2020}]%
        {Rieder2020}
\bibfield{author}{\bibinfo{person}{Bernhard Rieder} {and}
  \bibinfo{person}{Jeanette Hofmann}.} \bibinfo{year}{2020}\natexlab{}.
\newblock \showarticletitle{Towards platform observability}.
\newblock \bibinfo{journal}{\emph{Internet Policy Review}} \bibinfo{volume}{9},
  \bibinfo{number}{4} (\bibinfo{date}{Dec.} \bibinfo{year}{2020}).
\newblock
\urldef\tempurl%
\url{https://doi.org/10.14763/2020.4.1535}
\showDOI{\tempurl}


\bibitem[\protect\citeauthoryear{Ryoo and Moon}{Ryoo and Moon}{2014}]%
        {ryoo2014-inferring-twitter-location}
\bibfield{author}{\bibinfo{person}{KyoungMin Ryoo} {and} \bibinfo{person}{Sue
  Moon}.} \bibinfo{year}{2014}\natexlab{}.
\newblock \showarticletitle{Inferring Twitter User Locations with 10 Km
  Accuracy}. In \bibinfo{booktitle}{\emph{Proceedings of the 23rd International
  Conference on World Wide Web}} (Seoul, Korea) \emph{(\bibinfo{series}{WWW '14
  Companion})}. \bibinfo{publisher}{Association for Computing Machinery},
  \bibinfo{address}{New York, NY, USA}, \bibinfo{pages}{643–648}.
\newblock
\showISBNx{9781450327459}
\urldef\tempurl%
\url{https://doi.org/10.1145/2567948.2579236}
\showDOI{\tempurl}


\bibitem[\protect\citeauthoryear{Safi~Samghabadi, L{\'o}pez~Monroy, and
  Solorio}{Safi~Samghabadi et~al\mbox{.}}{2020}]%
        {safi-samghabadi-etal-2020-detecting}
\bibfield{author}{\bibinfo{person}{Niloofar Safi~Samghabadi},
  \bibinfo{person}{Adri{\'a}n~Pastor L{\'o}pez~Monroy}, {and}
  \bibinfo{person}{Thamar Solorio}.} \bibinfo{year}{2020}\natexlab{}.
\newblock \showarticletitle{Detecting Early Signs of Cyberbullying in Social
  Media}. In \bibinfo{booktitle}{\emph{Proceedings of the Second Workshop on
  Trolling, Aggression and Cyberbullying}}. \bibinfo{publisher}{European
  Language Resources Association (ELRA)}, \bibinfo{address}{Marseille, France},
  \bibinfo{pages}{144--149}.
\newblock
\showISBNx{979-10-95546-56-6}
\urldef\tempurl%
\url{https://aclanthology.org/2020.trac-1.23}
\showURL{%
\tempurl}


\bibitem[\protect\citeauthoryear{Schrepel}{Schrepel}{2021}]%
        {schrepel2021computational}
\bibfield{author}{\bibinfo{person}{Thibault Schrepel}.}
  \bibinfo{year}{2021}\natexlab{}.
\newblock \showarticletitle{Computational antitrust: An introduction and
  research agenda}.
\newblock \bibinfo{journal}{\emph{Computational Antitrust}}
  (\bibinfo{year}{2021}).
\newblock


\bibitem[\protect\citeauthoryear{Sigurbergsson and Derczynski}{Sigurbergsson
  and Derczynski}{2020}]%
        {sigurbergsson-derczynski-2020-offensive}
\bibfield{author}{\bibinfo{person}{Gudbjartur~Ingi Sigurbergsson} {and}
  \bibinfo{person}{Leon Derczynski}.} \bibinfo{year}{2020}\natexlab{}.
\newblock \showarticletitle{Offensive Language and Hate Speech Detection for
  {D}anish}. In \bibinfo{booktitle}{\emph{Proceedings of the 12th Language
  Resources and Evaluation Conference}}. \bibinfo{publisher}{European Language
  Resources Association}, \bibinfo{address}{Marseille, France},
  \bibinfo{pages}{3498--3508}.
\newblock
\showISBNx{979-10-95546-34-4}
\urldef\tempurl%
\url{https://aclanthology.org/2020.lrec-1.430}
\showURL{%
\tempurl}


\bibitem[\protect\citeauthoryear{Timberg}{Timberg}{2021}]%
        {fbsso}
\bibfield{author}{\bibinfo{person}{Craig Timberg}.}
  \bibinfo{year}{2021}\natexlab{}.
\newblock \bibinfo{title}{Facebook made big mistake in data it provided to
  researchers, undermining academic work}.
\newblock
\newblock
\urldef\tempurl%
\url{https://www.washingtonpost.com/technology/2021/09/10/facebook-error-data-social-scientists/}
\showURL{%
\tempurl}


\bibitem[\protect\citeauthoryear{Wu, Pedersen, and Salehi}{Wu
  et~al\mbox{.}}{2019}]%
        {10.1145/3359321}
\bibfield{author}{\bibinfo{person}{Eva~Yiwei Wu}, \bibinfo{person}{Emily
  Pedersen}, {and} \bibinfo{person}{Niloufar Salehi}.}
  \bibinfo{year}{2019}\natexlab{}.
\newblock \showarticletitle{Agent, Gatekeeper, Drug Dealer: How Content
  Creators Craft Algorithmic Personas}.
\newblock \bibinfo{journal}{\emph{Proc. ACM Hum.-Comput. Interact.}}
  \bibinfo{volume}{3}, \bibinfo{number}{CSCW}, Article \bibinfo{articleno}{219}
  (\bibinfo{date}{Nov.} \bibinfo{year}{2019}), \bibinfo{numpages}{27}~pages.
\newblock
\urldef\tempurl%
\url{https://doi.org/10.1145/3359321}
\showDOI{\tempurl}


\bibitem[\protect\citeauthoryear{Zhang, Hu, Zhang, and Liu}{Zhang
  et~al\mbox{.}}{2021}]%
        {Zhang_Hu_Zhang_Liu_2021}
\bibfield{author}{\bibinfo{person}{Jinxue Zhang}, \bibinfo{person}{Xia Hu},
  \bibinfo{person}{Yanchao Zhang}, {and} \bibinfo{person}{Huan Liu}.}
  \bibinfo{year}{2021}\natexlab{}.
\newblock \showarticletitle{Your Age Is No Secret: Inferring Microbloggers’
  Ages via Content and Interaction Analysis}.
\newblock \bibinfo{journal}{\emph{Proceedings of the International AAAI
  Conference on Web and Social Media}} \bibinfo{volume}{10},
  \bibinfo{number}{1} (\bibinfo{date}{Aug.} \bibinfo{year}{2021}),
  \bibinfo{pages}{476--485}.
\newblock
\urldef\tempurl%
\url{https://ojs.aaai.org/index.php/ICWSM/article/view/14731}
\showURL{%
\tempurl}


\end{thebibliography}

\end{document}